\documentclass[numberedappendix,twocolappendix,apj]{emulateapj}
\usepackage{ulem}
\usepackage{amsmath}
\usepackage{graphicx}
\usepackage{subfigure}
\usepackage{multirow}
\usepackage{color}
\shorttitle{Cosmic Rays in NGC 253 \& NGC 1068}
\shortauthors{Yoast-Hull, Gallagher, Zweibel \& Everett}
\begin{document}
\title{Active Galactic Nuclei, Neutrinos, and Interacting Cosmic Rays In NGC 253 \& NGC 1068}

\author{Tova M. Yoast-Hull$^{1,2}$, J. S. Gallagher III$^3$, Ellen G. Zweibel$^{1,2,3}$, and John E. Everett$^{4,5}$}

\affil{$^1$Department of Physics, University of Wisconsin-Madison, WI, USA; email: {\tt yoasthull@wisc.edu}}
\affil{$^2$Center for Magnetic Self-Organization in Laboratory and Astrophysical Plasmas, University of Wisconsin-Madison, WI, USA}
\affil{$^3$Department of Astronomy, University of Wisconsin-Madison, WI, USA}
\affil{$^4$Center for Interdisciplinary Exploration and Research in Astrophysics, Northwestern University, IL, USA}
\affil{$^5$Department of Physics \& Astronomy, Northwestern University, IL, USA}


\begin{abstract}
The galaxies M82, NGC 253, NGC 1068, and NGC 4945 have been detected in $\gamma$-rays by \textit{Fermi}.  Previously, we developed and tested a model for cosmic ray interactions in the starburst galaxy M82.  Now, we aim to explore the differences between starburst and active galactic nuclei (AGN) environments by applying our self-consistent model to the starburst galaxy NGC 253 and the Seyfert galaxy NGC 1068.  Assuming constant cosmic-ray acceleration efficiency by supernova remnants with Milky-Way parameters, we calculate the cosmic-ray proton and primary and secondary electron/positron populations, predict the radio and $\gamma$-ray spectra, and compare with published measurements.  We find that our models easily fits the observed $\gamma$-ray spectrum for NGC 253 while constraining the cosmic ray source spectral index and acceleration efficiency.  However, we encountered difficultly modeling the observed radio data and constraining the speed of the galactic wind and the magnetic field strength, unless the gas mass is less than currently preferred values.  Additionally, our starburst model consistently underestimates the observed $\gamma$-ray flux and overestimates the radio flux for NGC 1068; these issues would be resolved if the AGN is the primary source of $\gamma$-rays.  We discuss the implications of these results and make predictions for the neutrino fluxes for both galaxies.
\end{abstract}


\section{Introduction}

It is generally agreed that low-frequency (radio) emission from relativistic jets in active galactic nuclei (AGN) is synchrotron emission from nonthermal electrons.  However, it is still debated whether the high-frequency (X-ray \& $\gamma$-ray) emission is due solely to Compton scattering by the electrons responsible for the low-frequency synchrotron emission or due to hadronic processes resulting from protons being co-accelerated with the electrons \citep[e.g.,][]{Schlick02b}.  If hadronic models are valid, AGN would be an expected source of neutrinos, as the proton interactions responsible for the production of neutral pions which result in $\gamma$-rays also produce charged pions, which result in secondary electrons, positrons, and neutrinos \citep[e.g.,][]{Anchordoqui08}.  Can we expect AGN to be detectable neutrino sources at the level that starburst systems of similar $\gamma$-ray luminosity are expected to be?  Observations by the space-based $\gamma$-ray telescope \textit{Fermi} and future observations by the hard X-ray space telescope NuSTAR may help to further resolve this question.

\setcounter{footnote}{5}

Building on the approach of previous models \citep[e.g.][]{Torres04,Thompson07,Lacki10,Lacki11,Paglione12}, we developed a semi-analytic model for cosmic ray interactions for the starburst galaxy M82 \citep[][hereafter YEGZ]{YoastHull13}.  In developing the model, we emphasized the important role of galactic winds while limiting the number of free parameters and keeping our models observationally informed.  Now, we aim to determine whether the model is applicable to other starburst systems and explore benefits and limitations of such models due to the presence of AGN.  The starburst galaxy NGC 253 has been detected in $\gamma$-rays by both space-based (\textit{Fermi}) and ground-based (HESS) telescopes and possibly contains an active central supermassive black hole (SMBH) \citep{Mitsuishi11}.  Seyfert 2 galaxies NGC 1068 and NGC 4945 have also been detected with \textit{Fermi}.  Though NGC 1068 ($D = 14.4$ Mpc; NASA/IPAC Extragalactic Database)\footnote{The NASA/IPAC Extragalactic Database (NED) is operated by the Jet Propulsion Laboratory, California Institute of Technology, under contract with the National Aeronautics and Space Administration.} is more distant than the nearby NGC 253 \citep[$D = 3.9$ Mpc;][]{Karachentsev03} and NGC 4945 \citep[$D = 3.8$ Mpc;][]{Karachentsev07}, all three have $\gamma$-ray fluxes of the same order of magnitude \citep{Ackermann12}.  

In this paper, we determine whether our model is applicable to a combined starburst / AGN environment.  We select the starburst galaxy NGC 253 with a possible low-activity AGN and the high-activity Seyfert galaxy NGC 1068.  Applying our model to NGC 253 and NGC 1068, we seek the best-fit cosmic ray interaction model which accurately reproduces both radio and $\gamma$-ray spectra within the confines of the observed properties of the interstellar medium (ISM) and surrounding radiation field.  Additionally, we make a prediction of the neutrino fluxes from both galaxies.

\begin{center}
\begin{deluxetable*}{llclc}
%
\tablecaption{Observed Model Parameters}
\tablewidth{0pt}
\tablehead{
\colhead{Physical Parameters} & \colhead{NGC 253 Values Adopted} & \colhead{References} & \colhead{NGC 1068 Values Adopted} & \colhead{References}
}
\startdata
Distance & 3.9 Mpc & 1 & 14.4 Mpc & 7 \\
Central Molecular Zone (CMZ) Radius & 150 pc & 2 & 200 pc & 8 \\
Molecular Gas Mass & $1 - 3 \times 10^{8}$ $M_{\odot}$ & 3, 4 & $5 \times 10^{7}$ $M_{\odot}$ & 9 \\
Average ISM Density\tablenotemark{a} & 500 - 1400 cm$^{-3}$ & & 250 cm$^{-3}$ & \\
IR Luminosity & $3\times 10^{10}$ $L_{\odot}$ & 5 & $1.5 \times 10^{11}$ $L_{\odot}$ & 10 \\
Radiation Field Energy Density\tablenotemark{a} & 500 - 2000 eV~cm$^{-3}$ & & $10^{4}$ eV~cm$^{-3}$ & \\
SN Explosion Rate & 0.1 yr$^{-1}$ & 6 & 0.07 yr$^{-1}$ & 11 \\
SN Explosion Energy\tablenotemark{b} & 10$^{51}$ ergs & & 10$^{51}$ ergs & \\
SN Energy Transferred to CR\tablenotemark{b} & 4 - 20\% & & 10\% & \\
Ratio of Primary Protons & 50 & & 50 & \\
~~~to Electrons (N$_{p}$/N$_{e}$) & & & & \\
Slope of Primary CR & 2.2/2.3 & & 2.0 & \\
~~~Source Function & & & & \\
\enddata
%
%
\tablenotetext{a}{Derived from above parameters}
\tablenotetext{b}{Excludes neutrino energy}
\tablerefs{
(1)~\cite{Karachentsev03}; (2)~\cite{Sakamoto11}; (3)~\cite{Weiss08}; (4)~\cite{Hailey08}; (5)~\cite{Telesco80}; (6)~\cite{Lenc06}; (7)~NASA/IPAC Extragalactic Database; (8)~\cite{Krips11}; (9)~\cite{Schinnerer00}; (10)~\cite{Bock00}; (11)~\cite{Storchi12};
}
\end{deluxetable*}
\end{center}

The next section describes the observed properties of NGC 253 and NGC 1068.  Section 3 details how the population of energetic particles was computed.  Section 4 contains the results of our model for NGC 253 and NGC 1068.  In Section 5, we compare and contrast the results for NGC 253 and NGC 1068 and discuss predictions for the neutrino flux and the resulting implications.  In Section 6, we present concluding remarks.

\section{Galaxy Properties and Observations}

\subsection{Properties of NGC 253}

NGC 253 is a giant, barred spiral galaxy with a central starburst region.  Tidal interactions are the cause of many starbursts in massive galaxies, and while there is no obvious companion or trigger for the starburst in NGC 253, kinematic studies suggest NGC 253 may have been involved in a past merger \citep{Davidge10}.  Models indicate that the current high level of nuclear star formation has been ongoing for at least 20-30 Myr \citep{Engelbracht98}.

The center of NGC 253 is host to several compact objects which have been observed over a variety of wavelengths \citep{Forbes00}.  The strongest non-thermal, compact radio source (TH2) has long been considered a possible AGN candidate \citep{Ulvestad97, Muller10}.  Additionally, X-ray emission from the nucleus may be powered by an obscured AGN or a combination of supernova remnants and fluorescent line emission from molecular clouds \citep{Mitsuishi11}.  However, new observations with NuSTAR and \textit{Chandra} are unable to rule out the possibility that these compact X-ray sources are ultraluminous X-ray sources (ULXs) and not a low-luminosity AGN \citep{Lehmer13}.

One of the other major features of NGC 253 is the galactic wind emanating from the nuclear region \citep[e.g.,][]{Ulrich78, Fabbiano84, Schulz92}.  This galactic wind has been detected in both H$\alpha$ and X-rays.  Observations of the position and size of the outflow at both wavelengths match up almost exactly \citep{Strickland00, Strickland04}.  Optical observations give an outflow velocity on the order of a few 100 km~s$^{-1}$ \citep[e.g.,][]{Ulrich78, Schulz92, Westmoquette11}.  Additionally, \cite{Westmoquette11} suggest that, like star formation, the wind is quasi-steady.

Due to the obscured nature of the starburst nucleus of NGC 253 \citep[e.g.,][]{Kornei09, Waller88}, determining the properties of the nuclear region has been difficult.  Radio observations yield supernova rates ranging from 0.03 - 0.3 yr$^{-1}$ \citep{Ulvestad97, vanBuren94}.  In this work, we adopt a rate of 0.1 yr$^{-1}$ based on \cite{Lenc06} (see Table 1 for additional parameters).  Additionally, estimates of the molecular gas mass range from $(0.4 - 4.2) \times 10^{8}$ $M_{\odot}$ \citep[][and references therein]{Hailey08}.  We adopt a value of $3 \times 10^{8}$ $M_{\odot}$ for the molecular gas mass for this paper, based on work by \cite{Weiss08}.  We adopted an ionized gas mass for NGC 253 by scaling from the ratio of ionized to molecular gas in M82, which gives a mass of $\sim 3 \times 10^{6}$ $M_{\odot}$.

\subsection{NGC 253 Radio \& $\gamma$-Ray Spectra}

NGC 253 is well observed in the radio for both extended emission and the central starburst core.  Observations by \cite{Carilli96} and \cite{Williams10} at 333 MHz and from 1 to 7 GHz show that the radio spectrum turns down only slightly in the core at low frequencies, unlike the radio spectrum for the core of M82, which is completely absorbed at low frequencies.  High frequency observations from \cite{Ricci06} at 18.5 and 22 GHz and from \cite{Peng96} at 94 GHz require a free-free emission contribution to the flux in addition to non-thermal synchrotron emission.  Observed radio and $\gamma$-ray spectra are presented with model spectra in Section 4.

NGC 253 has been observed in $\gamma$-rays at both GeV and TeV energies by \textit{Fermi} and HESS.  \cite{Ackermann12} present four data points and two upper limits over 0.2 to 200 GeV from analysis of 30 months of \textit{Fermi} data.  \cite{Abramowski12} present a combined analysis of five data points and one upper limit over 0.2 TeV to 30 TeV from HESS data with the \textit{Fermi} data and fit a single power-law to both data sets, resulting in a spectral index of $\Gamma = 2.34$.  \cite{Paglione12} present a more current reduction of the \textit{Fermi} data which we use for comparison with our models instead of the \textit{Fermi} data shown in \cite{Ackermann12}.

\subsection{Properties of NGC 1068}

NGC 1068 is the closest, brightest Seyfert galaxy.  The total bolometric luminosity of the galaxy is $L_{IR} = 3 \times 10^{11}$ $L_{\odot}$; half of which is accounted for by the circumnuclear disk (CND) and AGN \citep{Bock00}.  The galaxy supports a kiloparsec-scale, non-thermal radio jet \citep{Wilson83} and a compact radio nucleus \citep{Gallimore96a, Muxlow96}.  The complex chemical and kinematic properties of the nuclear region have been well-studied at multiple wavelengths \citep{Exposito11, Krips11, Storchi12}.

The circumnuclear starburst ring surrounding the 2.3 kpc stellar bar \citep{Scoville88, Thronson89} is illuminated by star formation in the past 10 - 40 Myr \citep{Davies98}.  The stellar population inside the CND, the inner $\sim$200 pc, developed in two bursts at 30 and 300 Myr ago \citep{Storchi12}.  \cite{Storchi12} give the star formation rate (SFR) for the inner galaxy as $\sim$ 10 $M_{\odot}$ yr$^{-1}$.  This gives an upper bound on the supernova rate of $\sim$0.07 yr$^{-1}$ for the nuclear region.  We convert from SFR to supernova rate with by analysis similar to \cite{Condon92}.

Both starburst regions are rich in molecular gas with masses of $\sim 7 \times 10^{8}$ $M_{\odot}$ for the circumnuclear ring and $\sim 5 \times 10^{7}$ $M_{\odot}$ for the CND \citep{Schinnerer00, Spinoglio12}.  CO, HCN, and HCO$^{+}$ observations show that the gas is moderately dense ($n(H_{2}) = 10^{3.5} ~ - ~ 10^{5.5}$ cm$^{-3}$) with a kinetic temperature of $>100$ K \citep{Kamenetzky11, Krips11, Hailey12}.

\subsection{NGC 1068 Radio \& $\gamma$-Ray Spectra}

As the nearest and brightest Seyfert galaxy, NGC 1068 has been extensively studied in the radio.  On the kiloparsec-scale, radio observations have mapped the star-forming spiral arms and the radio jets.  There are extensive observations specifically targeting the AGN on the parsec-scale \citep{Honig08}.  However, as even observations of the inner hundred parsecs have a dominant contribution from the AGN \citep{Gallimore96a}, specifically the radio jet and counter-jet, there are essentially no observations of the Central Molecular Zone (CMZ) alone.  Radio maps show little evidence of ongoing star-formation in the central $\sim$150 pc.

$\gamma$-ray data for NGC 1068 are more limited than for NGC 253.  While there are four data points and two upper limits at GeV energies, there is only a single upper limit for TeV energies from HESS \citep{Ackermann12}.  The \textit{Fermi} data range from 0.2 GeV to 300 GeV and the HESS upper limit is at 0.2 TeV.

\section{Theoretical Approach}

\subsection{Primary and Secondary Cosmic Rays}

Diffusion is thought to be of minor importance in star-forming environments with high gas densities and galactic winds.  As such, following the approach in YEGZ, a cosmic ray spectrum can be calculated from a single quantity including the source function (dependent on supernova rate) and the lifetime.  Because we only consider radiative and advective losses, our model is not applicable to spatially extended regions, such as the inner spiral arms in NGC 1068, where cosmic ray diffusion is important.  Thus, we apply our model specifically to the CMZs of galaxies.  Supernovae are the assumed drivers of cosmic ray acceleration, and so we adopt a power-law source function ($Q(E) = A E^{-p}$) such that
\begin{equation}
\int_{E_{\text{min}}}^{E_{\text{max}}} Q(E) E dE = \frac{\eta \nu_{\text{SN}} E_{51}}{V} ,
\end{equation}
where $\nu_{\text{SN}}$ is the supernova rate, $V$ is the volume of the starburst region, $\eta$ is the fraction of the supernova energy transferred to cosmic rays, and $E_{51} = 1$ is 10$^{51}$ ergs, the typical energy from a supernova explosion.  Thus, the spectrum for primary protons is given by
\begin{equation}
N(E) = \frac{(p-2)}{E_{\text{min}}^{-p+2}} ~ \frac{\eta \nu_{\text{SN}} E_{51}}{V} E^{-p} ~ \tau(E),
\end{equation}
where $E_{\text{min}}$ is the minimum cosmic ray energy, here taken to be $E_{\text{min}} = 0.1$ GeV.  $\tau(E)$ is the cosmic ray lifetime composed of an energy loss timescale (due to radiative and collisional losses) and an energy independent timescale:
\begin{equation}\label{tau}
\tau(E)^{-1} \equiv \tau_{\text{adv}}^{-1} + \tau_{\text{loss}}^{-1} = \left( \frac{H}{v_{\text{adv}}}  \right)^{-1} + \left( - \frac{E}{dE/dt} \right)^{-1},
\end{equation}
where $H$ is the scale height of the starburst region and $v_{\text{adv}}$ is the speed of the particles in the wind of the starburst region \citep[see also][]{LackiBeck13}.  Energy losses include ionization, pion production, bremsstrahlung, synchrotron emission, and the inverse Compton effect.

Pion production from proton-proton collisions is the dominant energy loss for cosmic ray protons above the 1.3 GeV threshold.  As the lifetimes of both charged and neutral pions are very short, pions quickly decay into secondary electrons and positrons, neutrinos, and $\gamma$-rays.  The primary proton source function can be used to calculate the source function for charged and neutral pions \citep{Kelner06}, and the pion source function can then be used to calculate the spectrum of secondary electrons and positrons \citep{Schlick02}.

One of the main assumptions of the model is that cosmic rays sample the mean density of the interstellar medium.  The density of the interstellar medium sampled by the cosmic rays is vital to the model as it affects pion production and thus both the $\gamma$-ray and radio spectra.  The majority of the ISM is hot, low density gas in which pockets of cold, very dense molecular gas and warm ionized gas can be found.  In \cite{Boettcher13}, we assumed a simple two phase model of cold, molecular gas and hot, ionized gas of negligible density.  We found that for a wide range of cosmic ray injection conditions, magnetic field properties, and wind speeds, the cosmic rays sample the mean density of the ISM to within a factor of two.  The effect of a third phase of dense, warm, ionized gas on cosmic ray sampling will be considered in future work.

From cosmic ray proton and primary and secondary electron/positron populations, we can calculate nonthermal radio synchrotron emission and $\gamma$-ray emission by neutral pion decay, bremsstrahlung, and inverse Compton (see YEGZ for further details) for comparison with observed data.  Primary variables include spectral index ($p$), cosmic ray acceleration efficiency ($\eta$), molecular gas mass ($M_{\text{mol}}$), magnetic field strength ($B$), wind (advection) speed ($v_{\text{adv}}$), ionized gas density ($n_{\text{ion}}$), and interstellar radiation field energy density ($U_{\text{rad}}$).

\subsection{Radio Spectrum Model}

In YEGZ, we found that observations of the starburst core in M82 showed a complete turn down of the radio emission at low frequencies.  We attributed this to severe free-free absorption in the core and adopted a model for the starburst core with a cylindrical wall of warm, ionized gas surrounding a lower density hot medium.  Because some observations at low frequencies did not have enough resolution to distinguish the core from the halo in M82, we also included a component for radio emission from the halo.  Observations for NGC 253 do not show the same severe free-free absorption at low frequencies as in M82.  However, there is significant free-free emission at high frequencies, as is also the case in M82.  Thus, we expect that at least some fraction of the radio synchrotron emission is still absorbed by ionized gas.  To account for this, we keep the thin shell geometry (with shell width $l/2$ and radius $l$) for the nucleus, but assume only a small fraction ($f_{\text{abs}}$) of the synchrotron emission is absorbed by the ionized gas.  Then, the radiative intensity of the absorbed synchrotron emission is given by
\begin{equation}\label{diffuse}
I_{\nu, \text{abs}} = j_{\text{hot}} r_{i} e^{- \kappa_{\text{WIM}} l/2} f_{\text{abs}},
\end{equation}
where $r_{i}$ is the radius of the hot, diffuse gas ($r_{\text{starburst}} = r_{i} + l/2$), $j_{\text{hot}}$ is the emission coefficient for the hot gas ($j_{\text{hot}} = j_{\nu}^{\text{synch}} + j_{\nu}^{\text{ff},\text{hot}} \approx j_{\nu}^{\text{synch}}$), and $\kappa_{\text{WIM}}$ is the absorption coefficient for the warm, ionized gas ($\kappa_{\text{WIM}} = \kappa_{\nu}^{\text{ff},\text{ion}}$).  The majority of the radio spectrum is unabsorbed synchrotron emission from the hot, diffuse gas:
\begin{equation}\label{abs}
I_{\nu, \text{hot}} = j_{\text{hot}} r_{i} (1 - f_{\text{abs}}).
\end{equation}
Additionally, we have some portion of the radio spectrum composed of free-free emission from the ionized gas clouds:
\begin{equation}\label{ionized1}
I_{\nu, \text{ion}} = \frac{j_{\text{WIM}}}{\kappa_{\text{WIM}}} \left( 1 - e^{- \kappa_{\text{WIM}} l} \right),
\end{equation}
where $j_{\text{WIM}}$ is the emission coefficient for the warm, ionized gas ($j_{\text{WIM}} = j_{\nu}^{\text{synch}} + j_{\nu}^{\text{ff},\text{ion}}$).  Thus, the total emergent intensity from the starburst region is
\begin{equation}
I_{\nu} = j_{\text{hot}} r_{i} (f_{\text{abs}} + (1 - f_{\text{abs}}) e^{- \kappa_{\text{WIM}} l / 2}) + \frac{j_{\text{WIM}}}{\kappa_{\text{WIM}}} \left( 1 - e^{- \kappa_{\text{WIM}} l} \right).
\end{equation}
For an absorption fraction of $f_{\text{abs}} = 1$, this reduces to
\begin{equation}
I_{\nu} = j_{\text{hot}} r_{i} e^{- \kappa_{\text{WIM}} l / 2} + \frac{j_{\text{WIM}}}{\kappa_{\text{WIM}}} \left( 1 - e^{- \kappa_{\text{WIM}} l} \right), \nonumber
\end{equation}
which is the expression for the radio emission that we used for modeling M82 (YEGZ).  Note that unlike in YEGZ, we do not include a disk/halo component for radio emission in this paper as radio observations of the disk are separable from the halo \citep[e.g.,][]{Heesen09a, Heesen09b}.

\subsection{Gamma-Ray Spectrum Model}

We anticipate that inverse Compton scattering is a vital process for producing $\gamma$-rays in NGC 253 and NGC 1068.  For inverse Compton scattering, the $\gamma$-ray source function is given by \citep{RL79}
\begin{equation}
q_{\gamma,\text{IC}}(E_{\gamma}) = \frac{3 c \sigma_{T}}{16 \pi} \int_{0}^{\infty} d\epsilon \frac{v(\epsilon)}{\epsilon} \int_{\gamma_{\text{min}}}^{\infty} d\gamma \frac{n_{e}(\gamma)}{\gamma^{2}} F(q, \Gamma) ,
\end{equation}
with \citep{Schlick02}
\begin{equation*}
\gamma_{\text{min}} = \frac{E_{\gamma}}{(2 m_{e} c^{2})} \left[ 1 + \left( 1 + \frac{m_{e}^{2}c^{4}}{\epsilon E_{\gamma}} \right)^{1/2} \right],
\end{equation*}
where $E_{\gamma}$ is the energy of the resulting $\gamma$-ray, $\epsilon$ is the energy of the incident photon, $\gamma$ is the energy of the electron.  Here, the combined cosmic ray electron/positron spectrum, $n_{e}(\gamma)$, is in units of cm$^{-3}$.  The function $F(q, \Gamma)$ is part of the Klein-Nishina cross section and is given by \citep{Blumenthal70}
\begin{equation*}
F(q, \Gamma) = 2q \text{ln}(q) + (1 + q - 2q^{2}) + \frac{\Gamma^{2}q^{2} (1 - q)}{2(1 + \Gamma q)},
\end{equation*}
where
\begin{equation*}
\Gamma = \frac{4 \epsilon \gamma}{(m_{e}c^{2})} \text{ and } q = \frac{E_{\gamma}}{\Gamma (\gamma m_{e}c^{2} - E_{\gamma})}.  
\end{equation*}
For the blackbody spectrum, $v(\epsilon)$, we use an isotropic, diluted, modified blackbody spectrum \citep[][and references therein]{Persic08}
\begin{equation}
v(\epsilon) = \frac{C_{\text{dil}}}{\pi^{2} \hbar^{3} c^{3}} \frac{\epsilon^{2}}{e^{\epsilon / k T_{d}} - 1} \left( \frac{\epsilon}{\epsilon_{0}} \right)^{\sigma = 1},
\end{equation}
where $C_{\text{dil}}$ is a spatial dilution factor (given by the normalization $U_{\text{rad}} = \int v(\epsilon) \epsilon d\epsilon$) and $\epsilon_{0}$ corresponds to $\nu = 2 \times 10^{12}$ Hz.

\section{Model Results}
\subsection{NGC 253 Results}


%
%
\begin{figure*}[t!]
\epsscale{1.15}
\plottwo{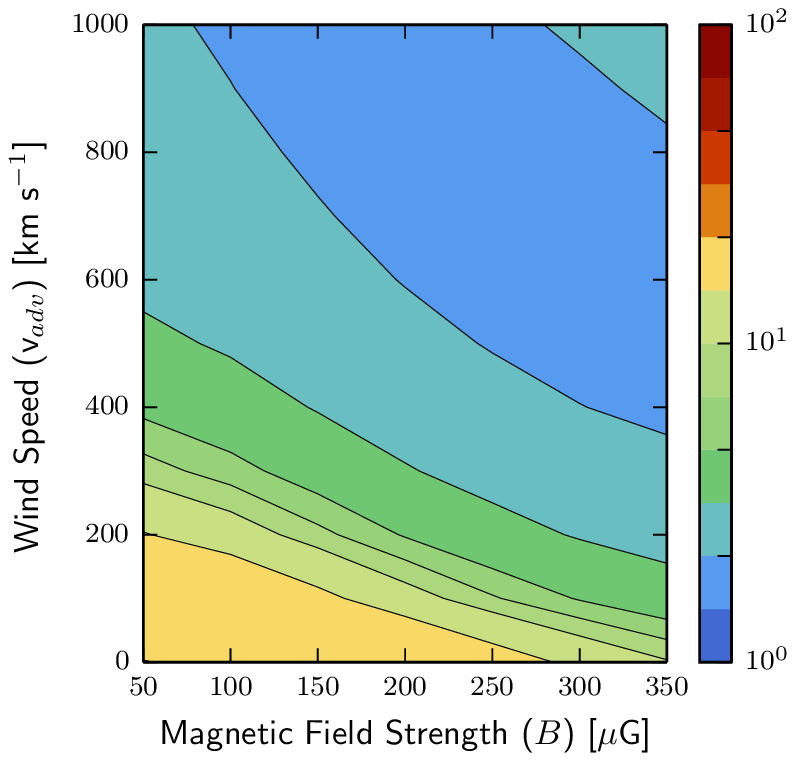}{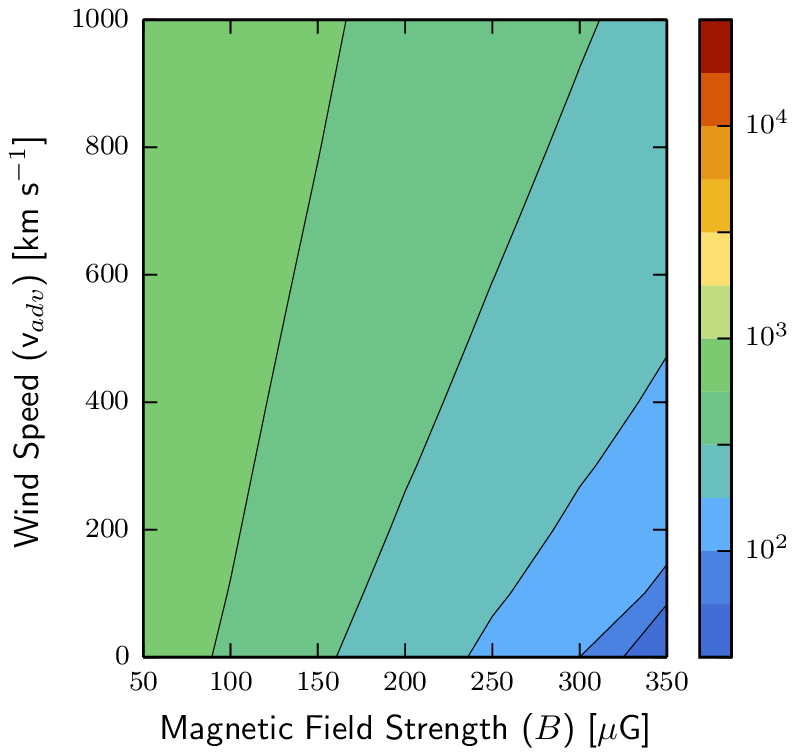}
\caption{Contour plots showing changes in reduced $\chi^{2}$ values for ranges of magnetic field strength ($B$) and advection (wind) speed ($v_{\text{adv}}$) for for fits to the $\gamma$-ray data (\textit{left}) and radio data (\textit{right}) for NGC 253.  While both plots show a degeneracy in magnetic field strength and wind speed, the trends are nearly orthogonal with the minimum of for each data set being at opposite ends of parameters space.  Model parameters are set at $p = 2.2$, $\eta = 0.04$, $M_{\text{mol}} = 3 \times 10^{8}$ $M_{\odot}$, $U_{\text{rad}} = 2000$ eV~cm$^{-3}$, $n_{\text{ion}} = 350$ cm$^{-3}$ (see Table 1 for additional parameters).}
\end{figure*}
\begin{figure*}[t!]
\epsscale{1.15}
\plottwo{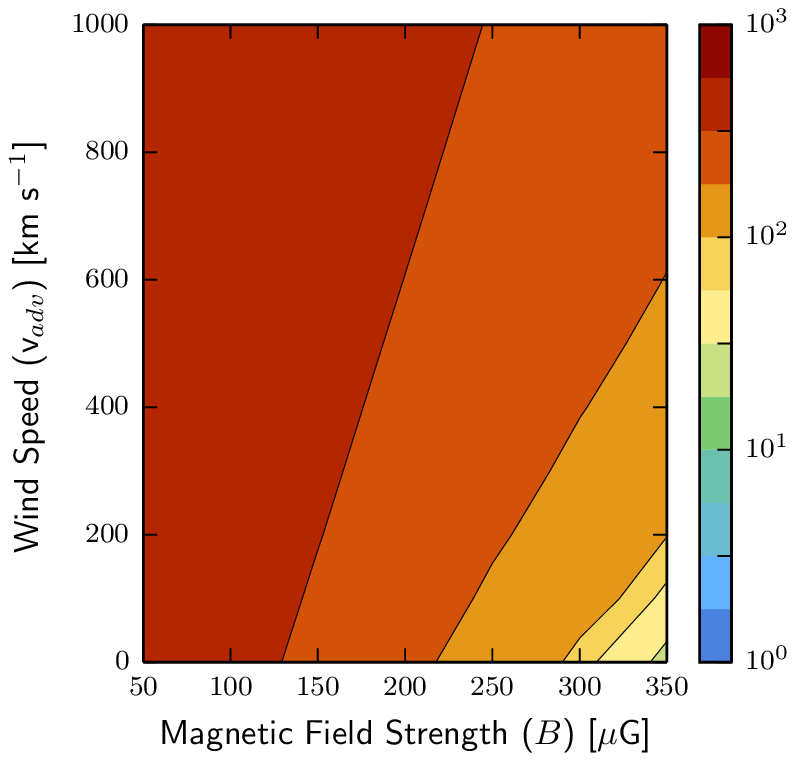}{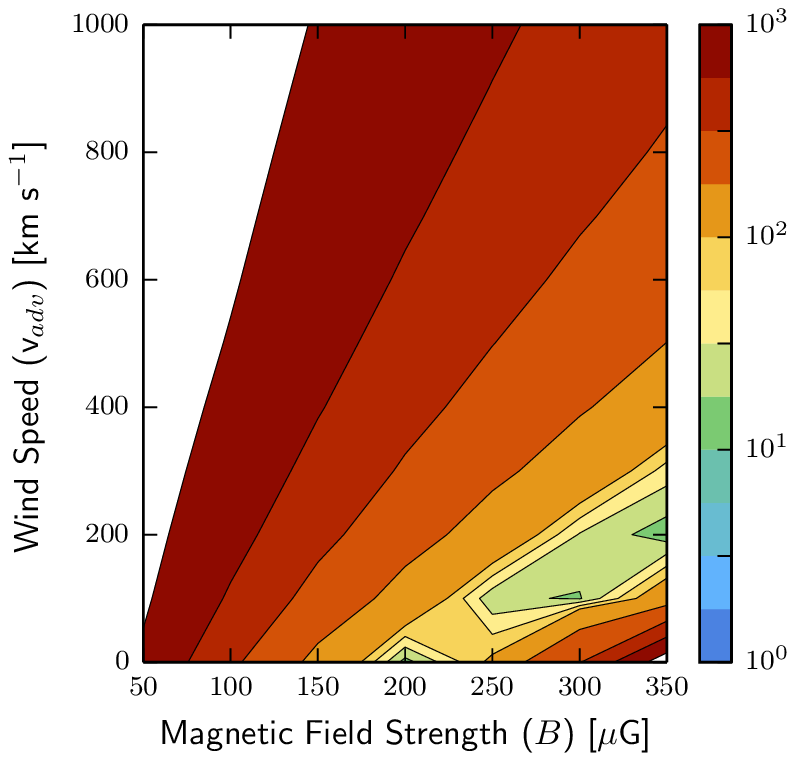}
\caption{Contour plots showing changes in reduced $\chi^{2}$ values for ranges of magnetic field strength ($B$) and advection (wind) speed ($v_{adv}$) for fits to the combined $\gamma$-ray and radio data for NGC 253.  We find that there is not a joint solution for the $\gamma$-rays and radio spectra and thus, the combined contour plot (\textit{left}) is dominated by the radio spectrum.  While the contour plot for the lower molecular mass (\textit{right}) is still dominated by the radio spectrum, there is a possible joint solution as denoted by the green contours.  Model parameters are set at (\textit{left}) $M_{\text{mol}} = 3 \times 10^{8}$ $M_{\odot}$, $U_{\text{rad}} = 2000$ eV~cm$^{-3}$ and (\textit{right}) $M_{\text{mol}} = 10^{8}$ $M_{\odot}$, $U_{\text{rad}} = 500$ eV~cm$^{-3}$ with $p = 2.2$, $\eta = 0.04$, $n_{\text{ion}} = 350$ cm$^{-3}$.}
\end{figure*}

As with our previous model for M82 (YEGZ), we test a range of parameters to find the best fits to both the radio and $\gamma$-ray spectra for the starburst core of NGC 253 (see Table 2).  We choose to vary parameters such as magnetic field strength ($B = 50 - 350$ $\mu$G) and radiation field energy density ($U_{\text{rad}} = 500 - 2000$ eV~cm$^{-3}$) as they are not well-constrained by observations.  We vary acceleration efficiency ($\eta = 0.04 - 0.2$) as it is equivalent to varying the supernova rate and test two spectral indices ($p = 2.2, 2.3$).  As the observed $\gamma$-ray spectrum for NGC 253 is steeper than that of M82, we choose larger spectral indices and do not test for $p = 2.1$.  We also test a variety of wind (advection) speeds ($v_{\text{adv}} = 0 - 1000$ km~s$^{-1}$) in order to study the no-wind case and wind speeds similar to those derived from optical and X-ray observations.  

\begin{center}
\begin{deluxetable}{ll}
%
\tablecaption{Varied Model Parameters for NGC 253}
\tablewidth{0pt}
\tablehead{
\colhead{Physical Parameters} & \colhead{Tested Range}
}
\startdata
Magnetic Field Strength ($B$) & 50 - 350 $\mu$G \\
Wind (Advection) Speed ($v_{\text{adv}}$) & 0 - 1000 km~s$^{-1}$ \\
Ionized Gas Density ($n_{\text{ion}}$) & 50 - 500 cm$^{-3}$ \\
Absorption Fraction ($f_{\text{abs}}$) & 0.1 - 1.0 \\
\enddata
\end{deluxetable}
\end{center}
\begin{figure*}[t!]
\epsscale{1.15}
\plottwo{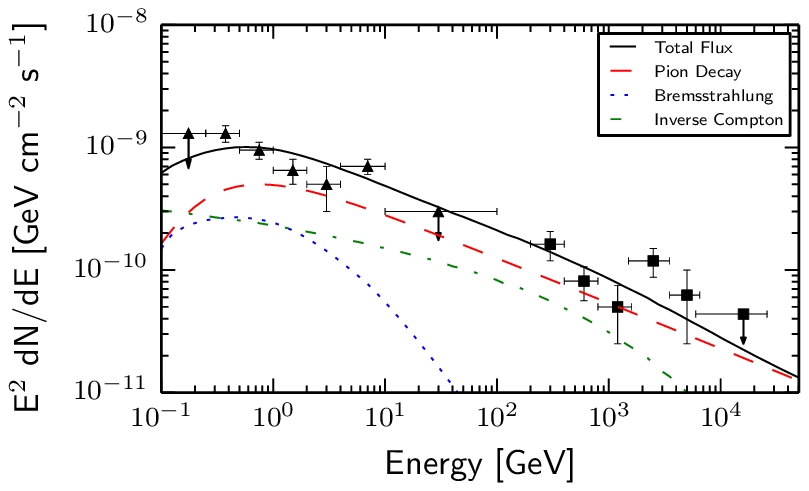}{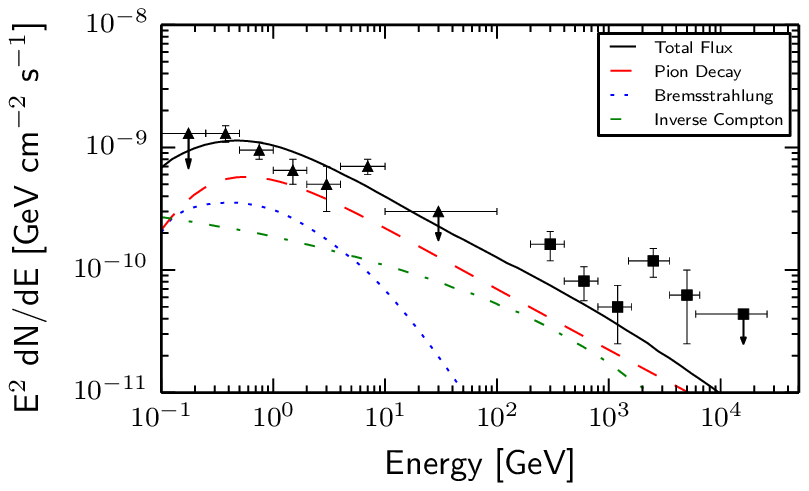}
\caption{Best-fit $\gamma$-ray spectra for NGC 253.  $\chi^{2}$ results for the $\gamma$-ray data eliminate nearly all combinations of spectral index and acceleration efficiency, except for $\eta = 0.04$ with $p = 2.2/2.3$.  However, as the model for spectral index $p = 2.3$ (\textit{right}) does not fit the HESS data, we exclude it also.  Model parameters are set at (\textit{left}) $p = 2.2$, $\eta = 0.04$, $U_{\text{rad}} = 2000$ eV~cm$^{-3}$, $n_{\text{ion}} = 350$ cm$^{-3}$, $v_{\text{adv}} = 600$ km~s$^{-1}$, $B = 300$ $\mu$G and (\textit{right}) $p = 2.3$, $\eta = 0.04$, $U_{\text{rad}} = 2000$ eV~cm$^{-3}$, $n_{\text{ion}} = 350$ cm$^{-3}$, $v_{\text{adv}} = 200$ km~s$^{-1}$, $B = 50$ $\mu$G with $M_{\text{mol}} = 3 \times 10^{8}$ $M_{\odot}$.  The solid lines represent the total $\gamma$-ray flux, the dashed lines represent the contribution from neutral pion decay, the dotted lines represent the contribution from bremsstrahlung, and the dot-dashed lines represent the contribution from inverse Compton.  $\gamma$-ray data include: \cite{Paglione12} (\textit{Fermi} - triangles), \cite{Abramowski12} (HESS - squares).  Data with downward arrows represent upper limits for both \textit{Fermi} and HESS data.}
\end{figure*}

In regards to the molecular gas mass, there is a wide a range of observed values and there is not yet a clear consensus on which is most accurate.  Varying the gas mass would have a similar (but not identical) effect to varying the acceleration efficiency or supernova rate.  In the case of the cosmic ray electrons, varying molecular gas mass affects the losses due to ionization and bremsstrahlung but not inverse Comtpon or synchrotron emission.  Though we adopt the value of $3 \times 10^{8}$ $M_{\odot}$, there is likely at least a factor of two uncertainty in this value.  As such, we also test models at $10^{8}$ $M_{\odot}$.  While we vary the molecular gas mass, we keep the ionized gas mass constant.  However, we do vary the ionized gas density ($n_{\text{ion}} = 50 - 500$ cm$^{-3}$), and thus the filling fraction of ionized gas, as it has a significant impact on free-free absorption and emission in the radio spectrum.

To distinguish between models, we use $\chi^{2}$ tests to compare to the observed data.  We can see from the contour plots of the $\chi^{2}$ values that the $\gamma$-ray and radio data constrain the value for the magnetic field strength and wind speed in different ways (see Figure 1).  There is a degeneracy for the magnetic field strength and wind speed for the radio data such that as the magnetic field strength increases, an increase in wind speed allows for a similarly good fit.  The reverse is true for the $\gamma$-rays.  In the radio spectrum, an increase in magnetic field strength results in an increase in synchrotron emission such that the number of electrons must be reduced (see YEGZ for further explanation).  However, in the $\gamma$-ray spectrum, an increase in the magnetic field strength leads to a reduction of the inverse Compton flux as fewer electrons are available as a result of the shorter energy loss lifetimes.  Thus, the wind speed must be reduced to allow for more electrons to be available to produce inverse Compton emission.

Because the results are essentially non-complementary, we were not able to find a joint solution (a best-fit model for both the $\gamma$-ray and radio spectra based on a common set of parameters) for the radio and $\gamma$-ray data assuming $3 \times 10^{8}$ $M_{\odot}$ for the molecular mass.  This is highlighted in Figure 2, where the combined reduced $\chi^{2}$ result is dominated by the radio data.  For the radio spectrum, optimal wind speeds are essentially the no wind case ($v_{\text{adv}}$) which is not supported by observations of NGC 253.  For the $\gamma$-ray spectrum, optimal wind speeds range from $v_{\text{adv}} = 400 - 1000$ km~s$^{-1}$, which is slightly higher than optical determinations of the wind speeds.  Additionally, while $\gamma$-ray models allow for a wide range of magnetic field strengths, radio models constrain the magnetic field strength to $300 - 350$ $\mu$G.

\begin{figure*}[t!]
\epsscale{1.15}
\plottwo{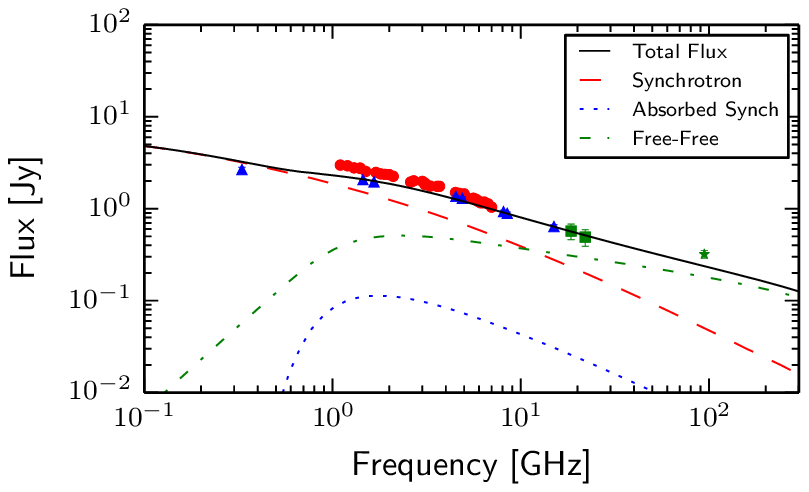}{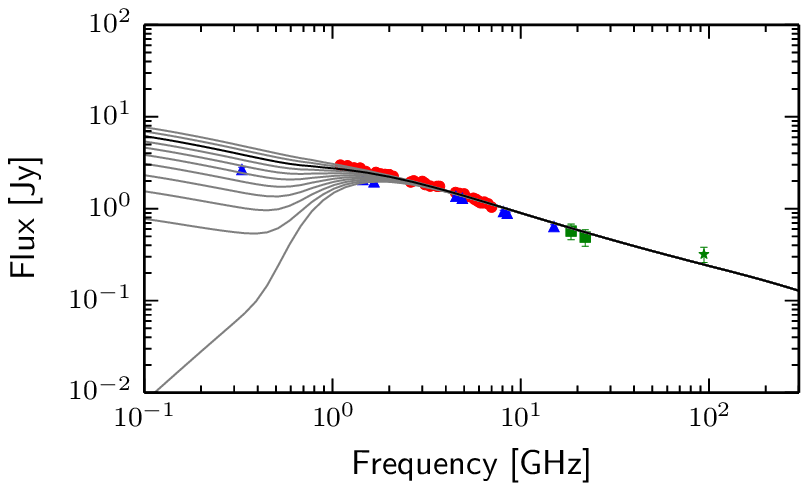}
\caption{Radio spectra for NGC 253.  The best-fit radio model is shown on the left.  The changes in the total radio spectrum as the fraction of absorbed synchrotron emission increases by varying the absorption fraction are shown on the right.  While we able to obtain a relatively good fit to the radio with the lower mass (\textit{right}), we were not able to achieve as good of a fit for the higher mass (\textit{left}) due to higher bremsstrahlung losses at low energies.  Model parameters are set at (\textit{left}) $p = 2.2$, $\eta = 0.04$, $U_{\text{rad}} = 2000$ eV~cm$^{-3}$, $n_{\text{ion}} = 350$ cm$^{-3}$, $v_{\text{adv}} = 0$ km~s$^{-1}$, $M_{\text{mol}} = 3 \times 10^{8}$ $M_{\odot}$ and (\textit{right}) $p = 2.2$, $\eta = 0.04$, $U_{\text{rad}} = 500$ eV~cm$^{-1}$, $n_{\text{ion}} = 350$ cm$^{-3}$, $v_{\text{adv}} = 200$ km~s$^{-1}$, $M_{\text{mol}} = 10^{8}$ $M_{\odot}$ with $B = 350$ $\mu$G.  The solid line denotes total radio flux, the dashed line represents the unabsorbed synchrotron radio emission in the hot, diffuse gas, the dotted line represents the free-free absorbed synchrotron radio emission in the hot, diffuse gas, and the dot-dashed line represents radio emission in the warm, ionized gas.  Radio data include \cite{Carilli96} (triangles), \cite{Williams10} (circles), \cite{Ricci06} (squares), and \cite{Peng96} (star).  Grey lines represent radio spectra with absorption fractions between 0.1 and 1.0 and the black line represents a radio spectrum with an absorption fraction of 0.2.}
\end{figure*}

These field strengths correspond to $\sim$2000 to $\sim$3000 eV~cm$^{-3}$.  Cosmic ray energy densities are nearly an order of magnitude below this, $\sim$200 to $\sim$300 eV~cm$^{-3}$ for $p = 2.2$.  While the magnetic field energy density lies at the high end of the range of the assumed radiation field energy densities, the cosmic ray energy densities are within a factor of two of the lower end of the range.  Taken at face value, these model results suggest that for the system the magnetic field and radiation field are in approximate equipartition while the cosmic rays are sub-equipartition.

In addition to limiting the possible magnetic field strength and wind speed values, the $\gamma$-ray data severely constrain the possible combinations of spectral index and acceleration efficiency.  While the radio data is fit equally well by all combinations of acceleration efficiency and spectral index, the $\gamma$-ray spectrum is easily overestimated because of the large radiation fields and high densities in starburst regions.  As such, the majority of the combinations of acceleration efficiency and spectral index do not result in a minimized (best-fit) $\chi^{2}$ value for the $\gamma$-ray spectrum.  So, to keep the combination of pion decay and inverse Compton emission in check, our model rules out acceleration efficiencies of $\eta = 0.1$ and  $\eta = 0.2$ for both spectral indices.  We also rule out the spectral index of $p = 2.3$ for all acceleration efficiencies, as the best-fit models underestimate the $\gamma$-ray spectrum at TeV energies (see Figure 3).  Thus, the only combination of acceleration efficiency and spectral index which both minimizes and fits the TeV energy data is $p = 2.2$, $\eta = 0.04$.  Equivalently, a larger acceleration efficiency pared with a lower supernova rate would have the same result, as we constrain their product.

The goodness of our $\chi^{2}$ fits also change slightly depending on the radiation field energy density.  The best-fit $\gamma$-ray spectrum occurs for an energy density of 2000 eV~cm$^{-3}$ for a spectral index of $p = 2.2$.  The resulting $\gamma$-ray spectrum is dominated by pion decay emission but still has a significant contribution from inverse Compton emission at TeV energies (see Figure 3).


Aside from the parameters discussed thus far, we also vary the fraction of synchrotron emission absorbed in ionized gas clouds (see Table 2).  Observations show that at low frequencies, the radio spectrum for NGC 253 levels off, unlike the radio spectrum for M82 which turns over at low frequencies.  In the case of M82, to fit the radio spectrum properly, we required that all of the synchrotron emission from the hot, diffuse medium was absorbed in ionized gas clouds such that $f_{\text{abs}} = 1.0$ (YEGZ).  However, as the radio spectrum for NGC 253 only flattens, we varied the absorption fraction to find what absorption fraction was required to accurately fit the spectrum.  We found that models with an absorption fraction of up to $f_{\text{abs}} = 0.2$ agreed with all compared radio data points (see Figure 4), including models with no absorption fraction ($f_{\text{abs}} = 0$).  The best-fit models were for an absorption fraction of $f_{\text{abs}} = 0.1$, which was selected for use for the remainder of our models.


In YEGZ, we found that while M82 was an electron calorimeter, it was, at best, only a partial ($\sim$50\%) proton calorimeter.  For NGC 253, we find that it is an electron calorimeter and that it can be a proton calorimeter, depending upon the selected parameters.  We based our conclusions on a comparison between energy-loss and advection timescales.  In NGC 253, we find that the average wind speeds for the best-fit $\gamma$-ray models are on the order of $v_{\text{adv}} = 400 - 1000$ km~s$^{-1}$.  The best-fit wind speed decreases depending on the assumed magnetic field strength.  Comparing timescales, we find that NGC 253 is a $\sim$50\% proton calorimeter.  However, the best-fits for the radio spectrum were for the no wind case.  In this case, NGC 253 would be a complete proton calorimeter \citep[cf.][]{Lacki11}.


While the $\gamma$-ray data were readily fit with our model, the radio spectrum presented much more of a challenge.  As previously stated, we were not able to find a joint solution for the $\gamma$-ray and radio models.  While our best-fit for the $\gamma$-ray spectrum had a reduced $\chi^{2}$ value of 2.0, the best-fit for the radio spectrum had a reduced $\chi^{2}$ value of 50.  In addition to the preferred wind speed being unphysical, the best-fit model does not match up well with the \cite{Williams10} data.  Although we did not explicitly vary molecular gas mass in our models, we did originally test a lower gas mass.  Models tested with a lower gas mass of $M_{\text{mol}} = 10^{8}$ $M_{\odot}$ resulted in significantly better radio fits (see Figures 2 and 4).  Our best-fit model with the lower mass has a reduced $\chi^{2}$ value of 12.  This reduction in $\chi^{2}$ with the lower mass is due to higher bremsstrahlung losses for higher masses.  Thus, for the lower mass, more electrons are available to produce synchrotron emission at the low frequencies instead of bremsstrahlung.

The discrepancy between fits for the different gas masses is, in large part, due to the limitations of the one-zone model.  While there are many computational advantages to assuming a simple single zone, there are also disadvantages.  In YEGZ, we noted that by assuming a single ionized gas density in our model, we were unable to fit both the high-frequency radio data and the low-frequency turn down.  In NGC 253, our one zone model has reached the calorimeter limit for the higher gas mass.  As such, while we are producing the same amount of secondary electrons and positrons at both masses, there are fewer electrons/positrons available to produce inverse Compton and synchrotron emission due to an increase in bremsstrahlung.

\subsection{NGC 1068 Results}


%
%
\begin{figure*}[t!]
\epsscale{1.15}
\plottwo{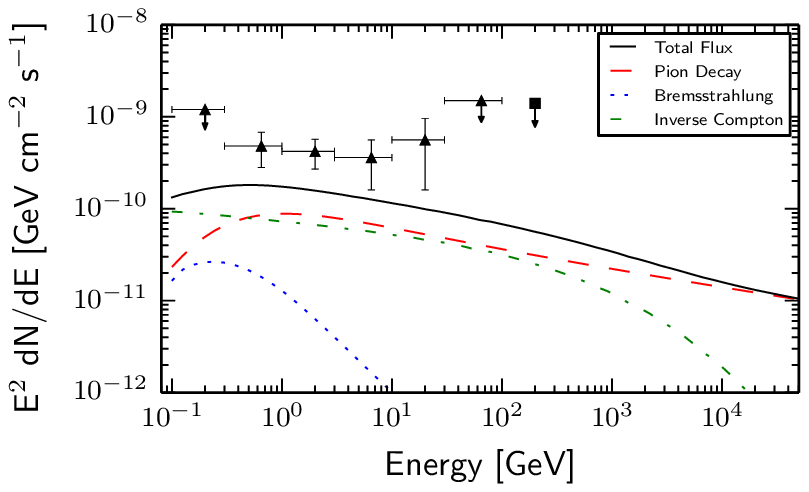}{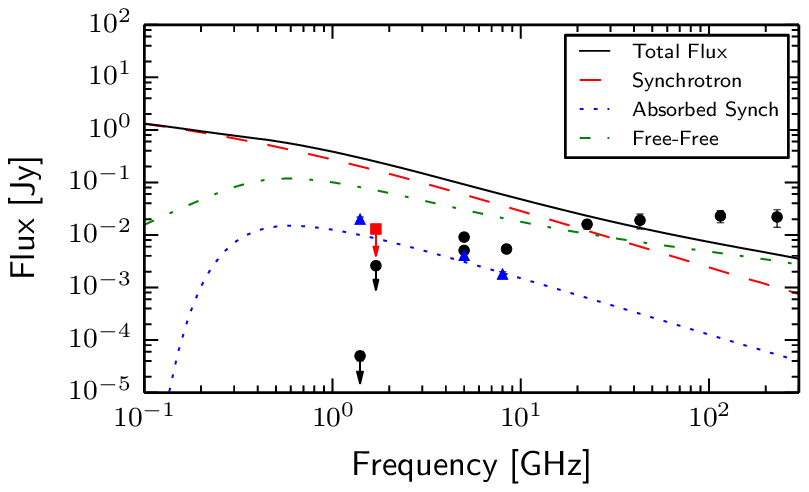}
\caption{$\gamma$-ray (\textit{left}) and radio (\textit{right}) spectra for NGC 1068.  While our models always underestimate the observed $\gamma$-ray flux, we also overestimate the radio flux.  Model parameters are set at $p = 2.0$, $\eta = 0.1$, $M_{\text{mol}} = 5 \times 10^{7}$ $M_{\odot}$, $U_{\text{rad}} = 10^{4}$ eV~cm$^{-3}$, $n_{\text{ion}} = 400$ cm$^{-3}$, $v_{\text{adv}} = 0$ km~s$^{-1}$, and $B = 200$ $\mu$G.  $\gamma$-ray data are represented as triangles for \textit{Fermi} data and squares for HESS data \citep{Ackermann12}.  Data with downward arrows represent upper limits for both \textit{Fermi} and HESS data.  Radio data are represented by blue triangles \citep[S2,][]{Gallimore04}, red square \citep[CMZ upper limit,][]{Gallimore96b}, and black circles \citep[S1,][]{Honig08}.}
\end{figure*}

The CMZ dust temperature is the key to modeling the $\gamma$-ray observations for NGC 1068 as it determines the radiation field that inverse Compton emission depends on.  Observations by \cite{Storchi12} show a blackbody spectrum with temperatures in the range of $700 \text{K} \leq T \leq 800 \text{K}$ for the inner CND.  When assuming a radiation field from dust with $T = 700$ K, the photon number is significantly decreased such that our models produce negligible inverse Compton $\gamma$-ray emission.  However, this dust temperature is attributed to the dusty torus of the AGN nucleus and likely does not dominate the larger, surrounding CMZ.  As such, we assume that the dust temperature in the CMZ is on par with the temperatures of molecular gas in the region, $\sim$100 K, and we use this to determine the radiation field spectrum.

As with NGC 253, we intended to test a variety of different sets of parameters with which to model NGC 1068.  However, we found NGC 1068 significantly harder to model than NGC 253.  The upper bound on the supernova rate produces a $\gamma$-ray spectrum that is lower by a factor of only a few (see Figure 5).  However, a lower bound on the supernova rate results in a $\gamma$-ray spectrum that is nearly two order of magnitudes lower than the observed data.  Because we were underestimating the $\gamma$-ray emission, we selected parameters to maximize the inverse Compton emission (a magnetic field strength of $B = 200$ $\mu$G and a radiation field energy density of $U_{\text{rad}} = 10^{4}$ eV~cm$^{-3}$) and the pion decay emission and bremsstrahlung (a wind speed of $v_{\text{adv}} = 0$ km~s$^{-1}$).  Even selecting parameters to augment the $\gamma$-ray emission, without invoking an extra source of cosmic rays, we were not able to produce a model which agrees with the \textit{Fermi} observations to better than a factor of a few.


Further complicating matters is the radio spectrum for NGC 1068.  While the galaxy has been extensively observed in the radio spectrum, the presence of a radio jet greatly overshadows any emission not originating from the AGN or its jets.  Ultimately, we chose to compare our radio models against a few different radio observations.  First, we plot our radio models against radio observations of the AGN core (S1) found in \cite{Honig08}.  Though the observations of the AGN are on the parsec-scale, they do highlight the fact that they are of a fundamentally different nature than the emission from star-forming regions.  We also plot radio observations of the region labeled S2 \citep{Gallimore04}, which could be the counter-jet as seen through the star-forming disk.

Additionally, we use the 18~cm radio map in \cite{Gallimore96b} to compare with our models.  While the outermost contour includes the spiral arms, the second contour should give us an upper limit on the radio continuum in the CMZ excluding the component from the AGN and the jets.  Taking the brightness level (of 74 mJy/beam) and multiplying by the area of the CMZ, we obtain an upper limit of $\sim$ 13 mJy (see red squared in Figure 5).

While nearly all of the energy of the cosmic ray electrons goes into inverse Compton emission, some small fraction of the energy is lost due to bremsstrahlung and synchrotron emission.  Assuming a moderate starburst magnetic field strength of 200~$\mu$G results in a significant amount of synchrotron emission.  As the upper limit (red square) that we plot is nearly an order of magnitude below the model, we must be significantly overestimating the synchrotron emission in the CMZ.  Thus, while assuming the upper limit on the supernova rate results in a $\gamma$-ray spectrum which comes within a factor of a few of the observed data, the radio spectrum disagrees significantly with observations (see Figure 5).

\section{Discussion}

\subsection{Starburst versus AGN Environments}

Previously, we developed and tested a model for cosmic ray interactions in starburst nuclei (YEGZ).  In testing our model against other nuclei and combined starburst / AGN environments, we have applied our updated model to two giant, barred spiral galaxies; specifically, we have modeled both NGC 253, with an unambiguous starburst nucleus, and NGC 1068, the archetypal Seyfert 2 galaxy.  The $\gamma$-ray spectrum for NGC 253 is well fit with a combination of emission from neutral pion decay, bremsstrahlung, and inverse Compton, requiring both hadronic and leptonic emission mechanisms.

\begin{figure*}[t!]
\epsscale{1.15}
\plottwo{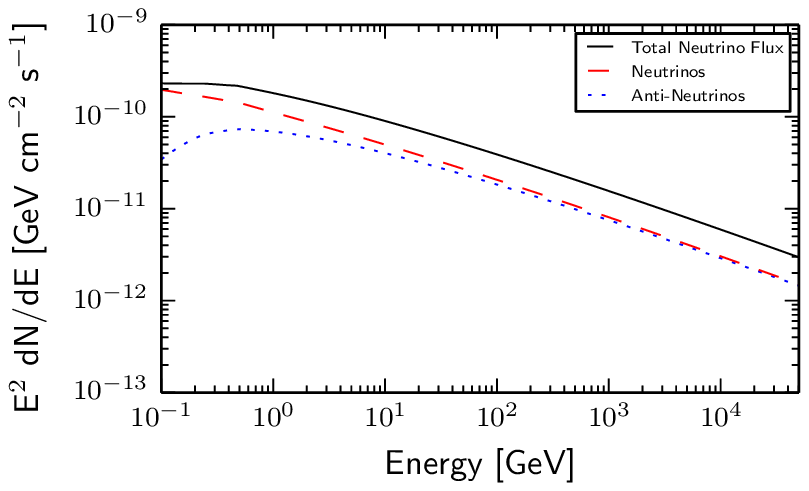}{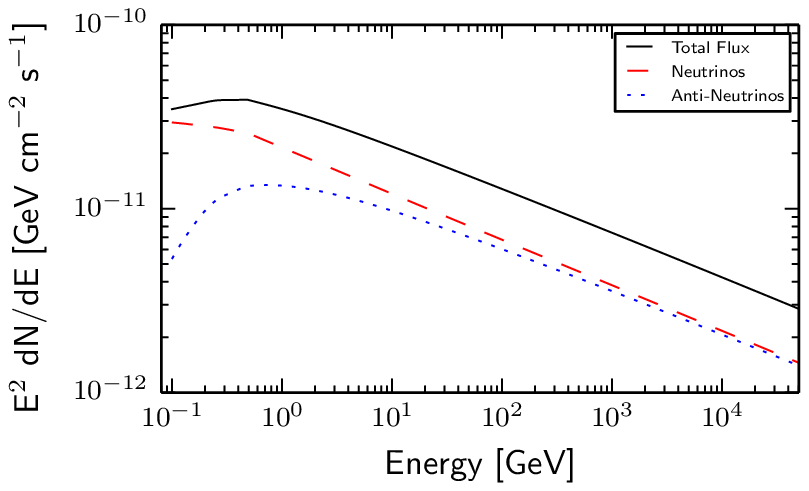}
\caption{Model neutrino spectra for NGC 253 (\textit{left}) and NGC 1068 (\textit{right}).  At TeV energies, we find that the predicted fluxes for NGC 253 and 1068 are comparable.  This is due to the fact that while there is nearly an order of magnitude different at GeV energies, NGC 1068 has a flatter spectrum than NGC 253.  Model parameters are set at $p = 2.2$, $\eta = 0.04$, $M_{\text{mol}} = 10^{8}$ $M_{\odot}$, $U_{\text{rad}} = 500$ eV~cm$^{-3}$, $n_{\text{ion}} = 400$ cm$^{-3}$, $B = 250$ $\mu$G, $v_{\text{adv}} = 100$ km~s$^{-1}$ (\textit{left}) and $p = 2.0$, $\eta = 0.1$, $M_{\text{mol}} = 5 \times 10^{7}$ $M_{\odot}$, $U_{\text{rad}} = 10^{4}$ eV~cm$^{-3}$, $n_{\text{ion}} = 400$ cm$^{-3}$, $B = 250$ $\mu$G, $v_{\text{adv}} = 0$ km~s$^{-1}$ with $\nu_{\text{SNR}} = 0.07$ yr$^{-1}$ (\textit{right}).  The solid black line denotes the total neutrino flux, the red dashed line represents the flux from muon neutrinos, the dotted blue line represents the flux from anti-neutrinos.}
\end{figure*}

While we find a joint solution for the $\gamma$-ray and radio spectra for the lower molecular gas mass for the NGC 253 CMZ, we were unable to do the same at a higher gas mass ($3 \times 10^{8}$ $M_{\odot}$).  \cite{Paglione12} find best-fit models for both the $\gamma$-ray and radio spectra for similarly large densities ($n_{ISM} = 1000$ cm$^{-3}$).  However, they do not include wind speed as a free parameter and fix their advection/diffusion timescale at 10 Myr (which is equivalent to a wind speed of $\sim$ 5 km~s$^{-1}$ for a half-thickness of 50 pc).  This would allow for a good fit to the radio spectrum at higher masses and their lower radiation field would also allow for a similarly good fit for the $\gamma$-ray spectrum.  Other models include those of \cite{Domingo05} who find joint solutions when assuming parameters closer to our lower gas mass and wind speeds between 300 and 600 km~s$^{-1}$.  We believe that the data support our choice of gas masses in the $1 - 3 \times 10^{8}$ $M_{\odot}$ range as well as wind speeds of $\geq 300$ km~s$^{-1}$ \citep[see][]{Weiss08,Westmoquette11} and that our models therefore contain astrophysically reasonable parameters.

Though located nearly four times further away, NGC 1068 has a $\gamma$-ray flux comparable to that of NGC 253.  In comparing our model with observations, we find that the inner starburst region in NGC 1068 does not supply a sufficient cosmic ray population to produce the observed $\gamma$-ray emission by itself.  This agrees with analysis of the supernova rate, total gas mass, and $\gamma$-ray luminosity by \cite{Lenain10} for NGC 1068 in comparison to other starburst galaxies.  While a scenario which boosts the cosmic ray population in the CMZ, such as a diffusion of cosmic rays from the circumnuclear ring into the CMZ, is possible, it is most likely that the high-energy emission results from interactions in the jet.  An AGN jet origin of the $\gamma$-ray emission in NGC 1068 agrees with calculations of \cite{Lenain10} who propose a leptonic external inverse Compton (EIC) model for the high-energy emission.

\subsection{Implications for Neutrino Fluxes}

While proton-proton interactions dominate as a cosmic ray energy loss mechanism in the ISM, proton-$\gamma$ interactions will be dominant in the jet due to the lack of molecular gas necessary for proton-proton interactions.  Proton-proton interactions result in both charged and neutral pions.  While neutral pions decay quickly into $\gamma$-rays, charged pions decay into muons and neutrinos followed by electrons and positrons and more neutrinos.  Thus, galaxies whose $\gamma$-ray spectra are dominated by or are significantly contributed to by pion decay will also be sources of neutrinos.  As proton-$\gamma$ interactions still result in pion production, AGN could also be a source for neutrinos as starbursts are, if $\gamma$-ray emission in the jet is hadronic.

While a detection of neutrinos from AGN would help to determine the origin of $\gamma$-ray emission, such a detection will also provide vital clues for starburst galaxies.  Observations of neutrinos provide direct information about the nature of the $\gamma$-ray spectrum and the cosmic ray proton population.  Better constraining the contribution of neutral pion decay to the $\gamma$-ray spectrum, also better constrains the contribution of bremsstrahlung and inverse Compton emission.  In the case of inverse Compton, this would be another diagnostic for determining the radiation field, which can be uncertain in such environments.

To make a neutrino flux prediction, we start with the source function for muon neutrinos which results from charged pion decay ($\pi^{+} \rightarrow \mu^{+} + \nu_{\mu}$ and $\mu^{+} \rightarrow e^{+} + \nu_{e} + \overline{\nu}_{\mu}$) depends on the source function for pions:
\begin{equation}
q_{\nu}(E_{\nu}) = \frac{1}{2 \eta} \int_{E_{\nu} / 2 \eta}^{\infty} \frac{q_{\pi}(E_{\pi})}{E_{\pi}},
\end{equation}
where $q_{\pi}(E_{\pi})$ is the source functions for pions (see YEGZ for details) and $\eta \equiv E_{\nu}^{\ast} / m_{\pi} = (m_{\pi}^{2} - m_{\mu}^{2}) / 2 m_{\pi}^{2}$ \citep{Stecker79}.

We found that the $\gamma$-ray spectra for NGC 253 had a significant contribution from pion decay.  While this results in a significant neutrino flux (see Figure 6), the steep spectrum of NGC 253 means that M82 is a more likely candidate for eventual detection of the energetic neutrinos for which IceCube has good sensitivity \citep{Halzen06, Abbasi11}.  Though all of our models for NGC 1068 underestimated the $\gamma$-ray emission, we find that the corresponding neutrino flux from the starburst region is comparable to that of NGC 253 at TeV energies (see Figure 6).  However, if the high-energy emission in the jet is hadronic, NGC 1068 may still be a detectable source of neutrinos, particularly if the flat $\gamma$-ray spectrum seen in Figure 5 continues to higher energies.  At energies below 1 GeV, models predict a significant difference in the spectra for inverse Compton and pion decay $\gamma$-ray emission \citep{Torres04}.  Thus, in the future, hard X-ray observations, such as those by NuSTAR, should be able to distinguish between leptonic and hadronic emission mechanisms for $\gamma$-ray production in such galaxies and determine whether these galaxies are likely to be detectable in neutrinos.


\section{Conclusions}

In this paper, we explore the differences between starburst galactic nuclei and AGN and their impact on cosmic ray interactions.  As a typical nuclear starburst environment, NGC 253 can be fit with our model that includes inverse Compton emission.  We find that for an assumed gas mass of $M_{\text{mol}} = 3 \times 10^{8}$ $M_{\odot}$, the $\gamma$-ray and radio spectra cannot uniquely determine the magnetic field strength or wind speed.  Additionally, the $\gamma$-ray spectrum also severely limits the possible combinations of spectral index (to $p = 2.2$) and acceleration efficiency (to $\eta = 0.04$) due to the powerful combination of effects due to inverse Compton and pion decay emission.

Past models for NGC 253 have assumed a fixed wind (advection) speed which presupposes calorimetry \citep[e.g.][]{Paglione96}.  Based on our results, a wind is a critical component for determining the degree to which a starburst system is a calorimeter.  While the best-fits to the $\gamma$-ray spectrum result in a $\sim$50\% proton calorimeter, fits to the radio spectrum result in a complete proton calorimeter.  Additionally, our fits result in systems below equipartition (in regards to the cosmic ray energy densities) with strong magnetic fields \citep[see also][]{Domingo05}.  The lack of a solution which fits the $\gamma$-ray and radio spectra together with a single set of parameters and the problems in fitting the radio spectrum highlight the limitations of this one-zone model.  We, however, can obtain a good fit to $\gamma$-rays and the radio by assuming a lower gas mass.  Understanding of the ISM conditions is vital to applying these models to starbursts.  As such, while the model works well for a late phase starburst such as M82, it does not as easily fit an early phase starburst such as NGC 253.

As an AGN environment containing a possible starburst region, NGC 1068 is not fit with our self-consistent starburst model.  While we find a few models that come close to predicting the observed $\gamma$-ray spectrum for NGC 1068, all of our models underestimate the high-energy emission and we are unable to find any models of the inner starburst region that agree with the observed data.  Thus, we find that the cosmic ray population from the starburst region alone is not enough to accurately fit the observed $\gamma$-ray emission and even those models in rough agreement with high-energy observations result in an overestimation of the radio spectrum.  We conclude that the observed $\gamma$-ray emission likely occurs in the AGN or its jets.  

Additionally, we find that starburst galaxies such as NGC 253 are potential sources for high energy neutrino detection.  However, AGN with high $L_{\gamma}$ such as NGC 1068 will not be luminous neutrino sources unless their high-energy emission is of hadronic origin.


\acknowledgements

This work was supported in part by NSF AST-0907837, NSF PHY-0821899 (to the Center for Magnetic Self-Organization in Laboratory and Astrophysical Plasmas), and NSF PHY-0969061 (to the IceCube Collaboration).  We thank Alberto Bolatto for conversations concerning NGC 253.  We also thank Francis Halzen for his help and support.


%
\end{document}